\documentclass{aastex631}

\usepackage{listings}
\submitjournal{ApJL}

\shorttitle{The Yangtze Stream}
\shortauthors{Yang et al.}
\graphicspath{{./}{figures/}}
\begin{document}

\title{A Dwarf Galaxy Debris Stream Associated with Palomar 1 and the Anticenter Stream}

\correspondingauthor{Jing-Kun Zhao}
\email{zjk@bao.ac.cn}

\author[0000-0001-7609-1947]{Yong Yang}
\affiliation{CAS Key Laboratory of Optical Astronomy, National Astronomical Observatories, Chinese Academy of Sciences, Beijing 100101, People's Republic of China}
\affiliation{School of Astronomy and Space Science, University of Chinese Academy of Sciences, Beijing 100049, People's Republic of China},
\author[0000-0003-2868-8276]{Jing-Kun Zhao}
\affiliation{CAS Key Laboratory of Optical Astronomy, National Astronomical Observatories, Chinese Academy of Sciences, Beijing 100101, People's Republic of China},
\author[0000-0002-5805-8112]{Xian-Hao Ye}
\affiliation{CAS Key Laboratory of Optical Astronomy, National Astronomical Observatories, Chinese Academy of Sciences, Beijing 100101, People's Republic of China}
\affiliation{School of Astronomy and Space Science, University of Chinese Academy of Sciences, Beijing 100049, People's Republic of China},
\author[0000-0002-8980-945X]{Gang Zhao}
\affiliation{CAS Key Laboratory of Optical Astronomy, National Astronomical Observatories, Chinese Academy of Sciences, Beijing 100101, People's Republic of China}
\affiliation{School of Astronomy and Space Science, University of Chinese Academy of Sciences, Beijing 100049, People's Republic of China},
\author[0000-0003-0173-6397]{Ke-Feng Tan}
\affiliation{CAS Key Laboratory of Optical Astronomy, National Astronomical Observatories, Chinese Academy of Sciences, Beijing 100101, People's Republic of China}



\begin{abstract}

We report the discovery of a new stream (dubbed as Yangtze) detected in $Gaia$ Data Release 3. The stream is at a heliocentric distance of $\sim$ 9.12 kpc and spans nearly 27$\degr$ by 1.9$\degr$ on sky. The colour–magnitude diagram of Yangtze indicates a stellar population of Age $\sim$ 11 Gyr and [M/H] $\sim$ -0.7 dex. It has a number density of about 5.5 stars degree$^{-2}$ along with a surface brightness of $\Sigma_G \simeq$ 34.9 mag arcsec$^{-2}$. The dynamics and metallicity estimate suggest that Yangtze may be closely related to Palomar 1 and the Anticenter stream. 

\end{abstract}



\section{Introduction} \label{sec:intro}

The Milky Way has proven to be full of substructures either embedded in the disk \citep[e.g.,][]{2009ApJ...692L.113Z,2018ApJ...868..105Z,2017ApJ...844..152L,2018Natur.561..360A,2021ApJ...922..105Y,2021ApJ...907L..16R,2021AJ....162..171Y,2021SCPMA..6439562Z} or hidden in the halo \citep[e.g.,][]{1994Natur.370..194I,2009ApJ...700L..61N,2010ApJ...714..229L,2016ASSL..420...87G,2018Natur.563...85H,2020ApJ...904...61Z,2021ApJ...914..123I,2022ApJ...935L..38Y,2022arXiv220410326M}. Identifying and studying them can help us better understand the current status as well as the past story of our Galaxy, which is especially true for stellar streams formed from disruption of satellite galaxies. This type of streams records accretion events that allow to reveal the formation history of the Milky Way. They have also maintained spatial clustering properties that make it possible to probe their kinematics and dynamics by tracing orbits \citep[e.g.,][]{2020ApJ...905..100C,2021ApJ...920...51M}. However, unlike globular-cluster-origin streams, the number of known dwarf galaxy streams is still limited so far (e.g., \citealp[Sagittarius,][]{1994Natur.370..194I}; \citealp[Orphan,][]{2006ApJ...645L..37G}; \citealp[Cetus,][]{2009ApJ...700L..61N}; \citealp[Tucana III,][]{2018ApJ...866...22L}; \citealp[LMS-1,][]{2020ApJ...898L..37Y}). Thus searching for this type of streams becomes of great importance and necessity.

In this letter, we report on the detection of a dwarf galaxy debris stream which we designate Yangtze. The stream is exposed by weighting stars in color-magnitude diagram (CMD) and proper motions (PMs) simultaneously using $Gaia$ Data Release 3 (DR3) \citep{2021A&A...649A...1G,2021A&A...649A...2L,2021A&A...649A...3R}. Section~\ref{sec:detection} describes the detecting strategy and Section~\ref{sec:stream} characterizes the stream. A conclusion is given in Section~\ref{sec:summary}. 

\section{A Stream Scanner} \label{sec:detection}

To search for streams of the Galactic halo, stars from $Gaia$ DR3 with galactic latitude $|b| > 15\degr$ are retrieved. In order to ensure good astrometric and photometric solutions, only stars with a renormalized unit weight error (RUWE) $<$ 1.4 and $|C^*| < 3\sigma_{C^*}$ are retained, where $C^*$ is the corrected BP and RP flux excess factor that was introduced by \citep{2021A&A...649A...3R} to identify sources for which the $G$–band photometry and BP and RP photometry are not consistent. 

A modified matched-filter technique has been adopted in \citet{2019ApJ...884..174G} and \citet{2022A&A...667A..37Y} to search for the tidal streams extended from globular clusters (GCs). The technique weights stars using their color differences from the cluster’s locus in CMD. These weights are further scaled based on stars’ departures from PMs of the cluster's orbit. Nevertheless, the limitation is that we must know the location and velocity of the progenitor such that we can integrate its orbit along which we assign weights to stars. In this work, we improve this method to make it applicable even if there is no prior knowledge about the stream progenitor. 

In CMD, we use an isochrone of old and metal-poor stellar population (e.g., Age = 13 Gyr and [M/H] = -2.2 dex) extracted from Padova database \citep{2012MNRAS.427..127B} as the filter. Individual stars are assigned weights based on their color differences from the isochrone, assuming a Gaussian error distribution:
\begin{equation}
	w_{\rm CMD} = \frac{1}{\sqrt{2\pi} \sigma_{color}} {\rm exp}
	\left[ -\frac{1}{2} \left(\frac{color - color_0}{\sigma_{color}} \right)^2  \right]  .
\end{equation}
Here $color$ and $\sigma_{color}$ denote $BP - RP$ and corresponding errors. $\sigma_{color}$ is simply calculated through $\sqrt{\sigma^2_{BP} + \sigma^2_{RP}}$ where $\sigma_{BP}$ and $\sigma_{RP}$ are obtained with a propagation of flux errors (see CDS website\footnote{\url{https://vizier.u-strasbg.fr/viz-bin/VizieR-n?-source=METAnot&catid=1350&notid=63&-out=text}.}). $color_0$ is determined by the isochrone at a given $G$ magnitude of a star. All stars have been extinction-corrected using the \citet{1998ApJ...500..525S} maps as re-calibrated by \citet{2011ApJ...737..103S} with $RV$ = 3.1, assuming $A_{G}/A_{V} = 0.83627$, $A_{BP}/A_{V} = 1.08337$, $A_{RP}/A_{V} = 0.63439$\footnote{These extinction ratios are listed on the Padova model site \url{http://stev.oapd.inaf.it/cgi-bin/cmd}.}. Therefore, $w_{\rm CMD}$ is a function of distance modulus ($dm$) once the isochrone is selected. In terms of PMs, weights are computed as:
\begin{equation}
	w_{\rm PMs} = \frac{1}{2\pi \sigma_{\mu^*_{\alpha}}\sigma_{\mu_{\delta}}} {\rm exp}
	\left\lbrace -\frac{1}{2} \left[ \left( \frac{\mu^*_{\alpha} - \mu^*_{\alpha,0}}{\sigma_{\mu^*_{\alpha}}} \right)^2 +
	\left( \frac{\mu_{\delta} - \mu_{\delta,0}}{\sigma_{\mu_{\delta}}} \right)^2 \right] \right\rbrace .
\end{equation}
Here $\mu^*_{\alpha}$, $\mu_{\delta}$, $\sigma_{\mu^*_{\alpha}}$ and $\sigma_{\mu_{\delta}}$ are measured PMs and corresponding errors of stars. $\mu^*_{\alpha,0}$ and $\mu_{\delta,0}$ are two variables of $w_{\rm PMs}$ varying in PM space. 

Finally, stars weights are obtained by multiplying both $w = w_{\rm CMD} \times w_{\rm PMs}$, which is a function of $dm$, $\mu^*_{\alpha,0}$ and $\mu_{\delta,0}$ (i.e., $w$ = $w(dm, \mu^*_{\alpha,0}, \mu_{\delta,0})$). We calculate $w$ for all stars over a grid of the three variables: $dm$ from 11.5 to 17.5 mag with a step of 0.2 mag, $\mu^*_{\alpha,0}$ from -15 to 15 mas/yr and $\mu_{\delta,0}$ from -20 to 10 mas/yr, both with a step of 1 mas/yr\footnote{These ranges and steps might be adjusted depending on specific goals and computational abilities.}.  At each grid point, stars' weights are summed in sky pixels to expose structures. We refer to this method as \texttt{StreamScanner}\footnote{Its \texttt{Python} code can be found at \url{https://doi.org/10.5281/zenodo.7604477}.} later because what it is doing is like scanning streams in CMD and PM space. 

Apparently, \texttt{StreamScanner} is a simple and direct way that relies on measurements from observation and does not need too complicated assumptions or models. We happen to find the signature of Yangtze in this work when \texttt{StreamScanner} is placed at $(dm, \mu^*_{\alpha,0}, \mu_{\delta,0})$ = (13.7, -1, 1). We then refine these parameters along with choice of the isochrone as elaborated in the next Section. 

\section{The Yangtze Stream} \label{sec:stream}

\begin{figure*}
	\includegraphics[width=\linewidth]{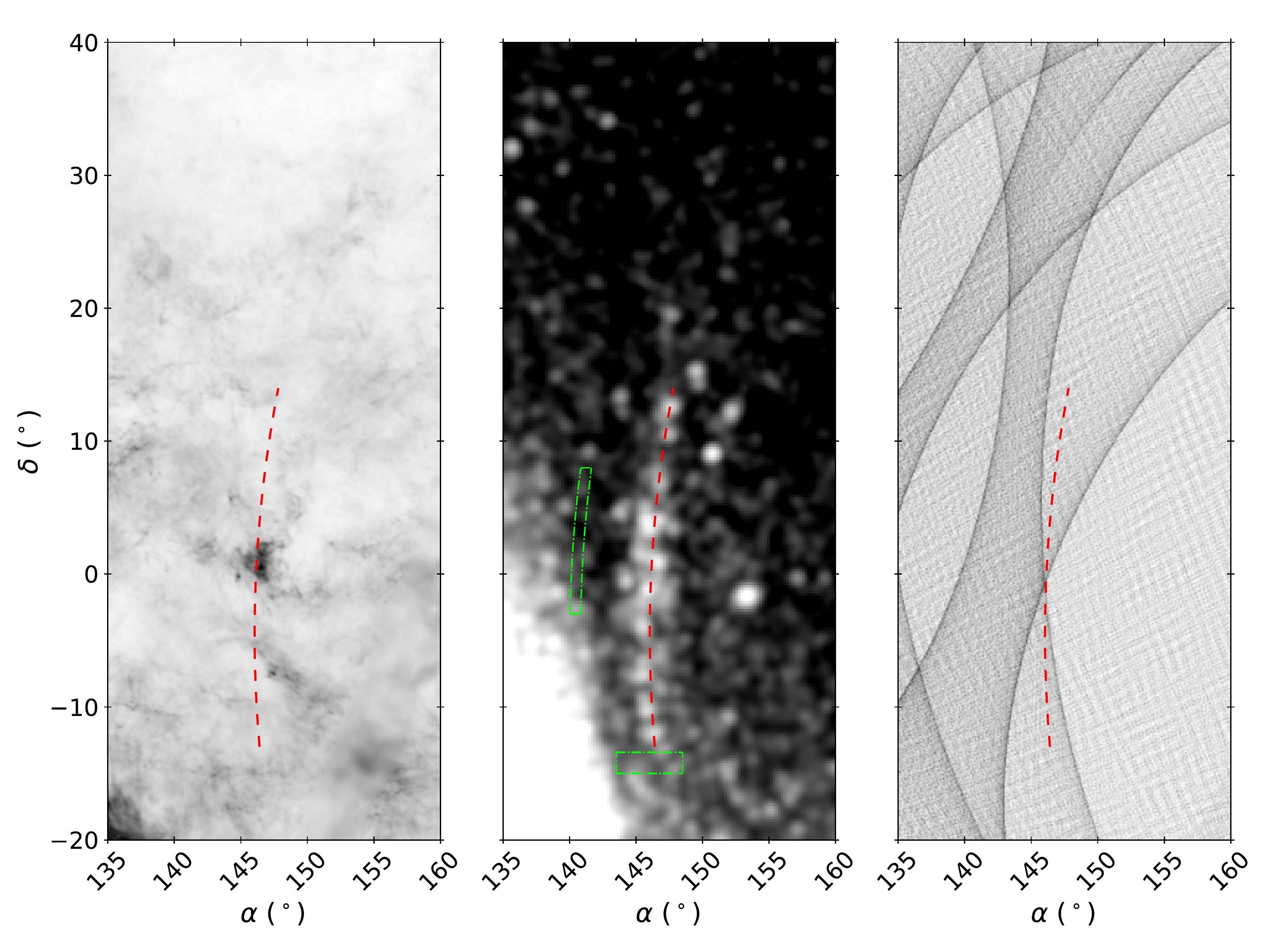}
	\caption{The left and right panels present the dust extinction map extracted from \citet{1998ApJ...500..525S} and $Gaia$'s scanning pattern covered by the DR3, respectively. The middle panel presents the weighted sky map for stars fainter than $G$ = 18 mag. The green areas are two moving masks used to generate the stream's lateral and longitudinal profiles in Figure~\ref{fig:profile}. The red dashed line of the three panels indicates the trajectory of the stream.
		\label{fig:ra_dec}}
\end{figure*}

A weighted sky map is obtained as shown in the middle panel of Figure~\ref{fig:ra_dec}. This is a coadded result of $w(14.8, -0.4, 0.8)$ and $w(14.8, -1, 1.4)$ using an isochrone with Age = 11 Gyr and [M/H] = -0.7 dex, where the isochrone and its $dm$ are determined by a CMD fit to the stream (see Section~\ref{subsec:cmd_pm}) and the choices of $(\mu^*_{\alpha,0}, \mu_{\delta,0})$ are based on Figure~\ref{fig:CMD_PMs}. The sky pixel width is 0.2$\degr$ and the coadded map is smoothed with a Gaussian kernel of $\sigma$ = 0.4$\degr$. The stretch is logarithmic, with brighter areas corresponding to higher weight regions. The white area in the bottom left corner is due to the proximity to the Galactic disk. Furthermore, only stars fainter than $G$ = 18 mag (corresponding to the stream's main sequence turn-off, see Figure~\ref{fig:CMD_PMs}) are shown here because random brighter field stars might introduce large noises due to their rather low photometric and astrometric uncertainties. On the left and right panels, we further plot the dust extinction map extracted from \citet{1998ApJ...500..525S} and $Gaia$'s scanning pattern covered by the DR3, respectively, with higher values represented by darker colors. The red dashed line of the three panels indicates the trajectory of Yangtze.

From the matched-filter map, the signature of Yangtze starts at $\delta$ = -13$\degr$ and ends at $\delta$ = 14$\degr$. Besides, there are several random noises appearing around the stream, which do not represent actual physical overdensities. According to the other two panels, we can verify that the stream does not follow any structures in the interstellar extinction and is not aligned with any features in $Gaia$’s scanning pattern. Although there is a small region at $\delta \sim$ 0$\degr$ where dust might be heavy, it does not resemble the stream pattern. To make sure that dereddening photometry with the map from \citet{1998ApJ...500..525S} is appropriate for this region, we compare the extinction to that extracted from 3D dust map \citep{2016ApJ...818..130B} by assuming the stream distance of 9.12 kpc (see CMD fitting in Section~\ref{subsec:cmd_pm}) and find that the typical difference in $A_V$ is only $\sim$ 0.04 mag, meaning that the 2D dust map we used from \citet{1998ApJ...500..525S} is accurate enough. The trajectory can be well described with a second-order polynomial:
\begin{equation}
	\alpha = 5.194 \times 10^{-3} \delta^2 + 4.685 \times 10^{-2} \delta + 146.141
	\label{equation3}
\end{equation}
with -13$\degr$ $<$ $\delta$ $<$ 14$\degr$. 

\begin{figure*}
	\includegraphics[width=\linewidth]{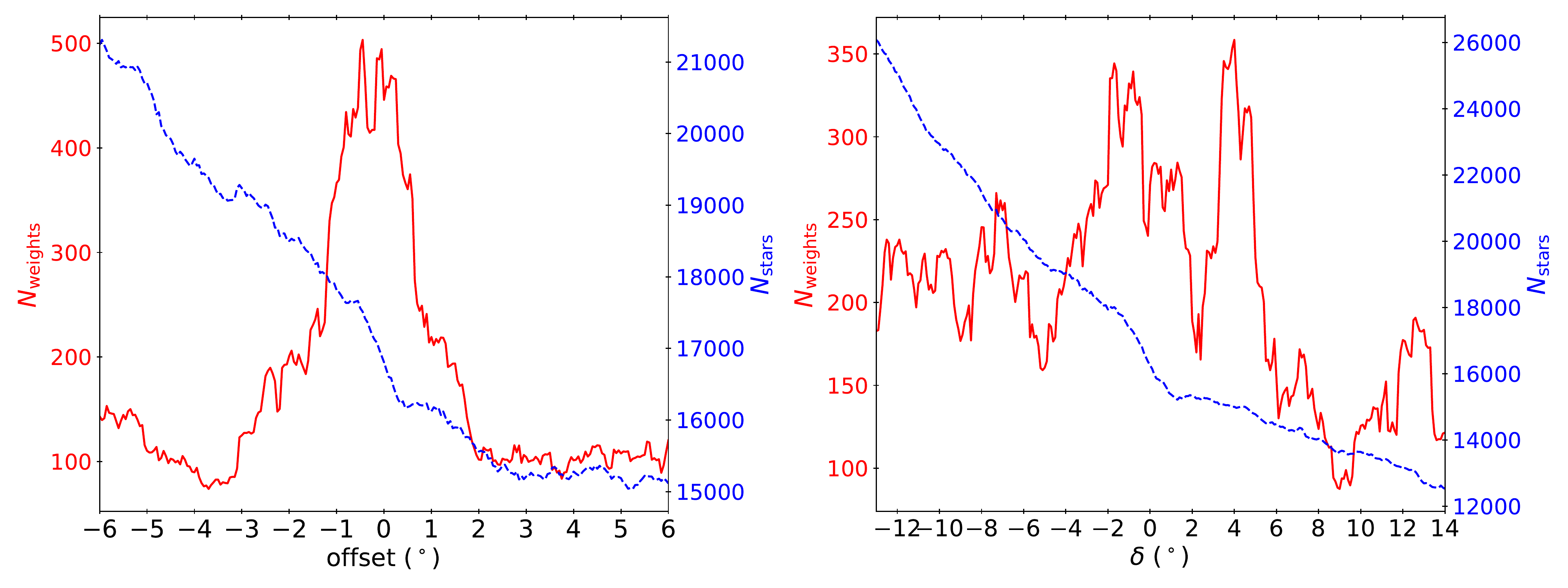}
	\caption{Left: the lateral distributions for weighted (red solid) and unweighted (blue dashed) numbers of stars fainter than $G$ = 18 mag, obtained by moving the left mask in Figure~\ref{fig:ra_dec} across the stream. Right: similar to the left panel but for the longitudinal profile obtained by moving the bottom mask along the stream.
		\label{fig:profile}}
\end{figure*}

To estimate the stream's width, we create a mask as shown with the left green window in Figure~\ref{fig:ra_dec}. Here $\delta$ is between -3$\degr$ and 8$\degr$ and this segment corresponds to the most prominent portion of the detected features. The boundaries on left and right sides are parallel to Equation~\ref{equation3} and separated by 0.8$\degr$ (the mask's width). We then move the mask across the stream from left to right and sum all weights of stars fainter than $G$ = 18 mag in the mask (again, to reduce large noises caused by bright stars), to create a one-dimensional stream profile as shown with the red solid line in the left panel of Figure~\ref{fig:profile}. This is directly analogous to the T statistic of \citet{2009ApJ...693.1118G}. From Figure~\ref{fig:profile}, the stream is almost enclosed within $-2\degr <$ offset $< 2\degr$, and its peak is 14.88$\sigma$ above the background noise outside this range. From the profile, we find an estimate of its width (full width at half maximum) to be $\sim 1.9\degr$. 

Furthermore, we create the lateral profile of unweighted stars in the same way and overplot it with the blue dashed line in left panel of Figure~\ref{fig:profile}. There is a gradient in the distribution of stars along offset, with more stars populating at lower offset (near the disk). It can be concluded that the stream signature is not caused by contamination of the Milky Way's field population. Otherwise, it is more likely to detect strong signals close to offset = -6$\degr$ side. 

Following the same way, the stream's longitudinal profile can be obtained using the bottom mask (5$\degr$ by 1.6$\degr$) in Figure~\ref{fig:ra_dec}, by moving it along the declination. As displayed in the right panel of Figure~\ref{fig:profile}, the red solid and blue dashed lines are presenting the same meaning as those of the left panel. From the profile of weights, several gaps are found at, for example, $\delta \sim$ -5$\degr$, 2$\degr$ and 9$\degr$, which also appear obvious in Figure~\ref{fig:ra_dec}. This kind of signature may suggest a result of non-continuous stripping \citep{2011ApJ...738...98G}, or collisions with massive perturbers in the Milky Way \citep{2019ApJ...880...38B}.

\subsection{CMD and PMs}\label{subsec:cmd_pm}

\begin{figure*}
	\includegraphics[width=\linewidth]{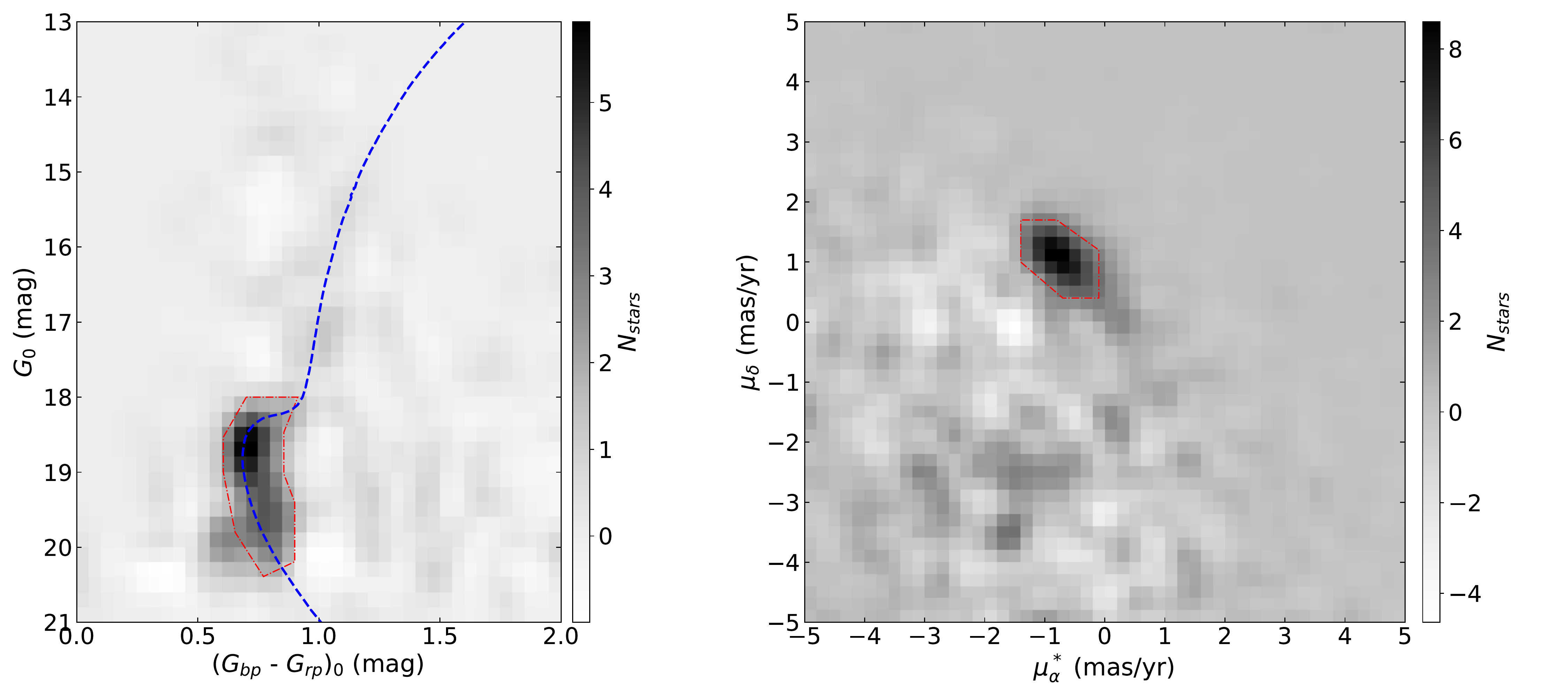}
	\caption{The left panel is a 2D histogram of stars in CMD with PMs selected and background subtracted. The blue dashed line represents the best-fit isochrone with Age = 11 Gyr and [M/H] = -0.7 dex at $dm$ = 14.8 mag. The right panel is a 2D histogram of PMs after CMD selection and background subtraction. Both of the diagrams are smoothed with a 2D Gaussian kernel of $\sigma$ = 1 pixel. The red polygons represent the CMD and PM selections applied to the stream and off-stream regions.
		\label{fig:CMD_PMs}}
\end{figure*}

We display a background-subtracted binned CMD for the stream in the left panel of Figure~\ref{fig:CMD_PMs}. The stream region is defined as the area around the trajectory (Equation~\ref{equation3}) $\pm 1\degr$ in $\alpha$ direction, given the derived width of 1.9$\degr$. The background is estimated through averaging two off-stream regions parallel to the stream obtained by moving the stream region along $\alpha$-axis by $\pm 3\degr$, to eliminate the effect of the gradient. Before the background subtraction, a PM selection is applied to both of the stream and off-stream regions as illustrated with the red polygon in the right panel of Figure~\ref{fig:CMD_PMs}, which corresponds to the stream's distribution in PM space (see below). We emphasize that this is a subtraction of star numbers, not weighted counts. 

The CMD bin size is 0.05 mag in color and 0.2 mag in $G$ magnitude. The diagram is smoothed with a 2D Gaussian kernel of $\sigma$ = 1 pixel. The blue dashed line represents the best-fit isochrone with Age = 11 Gyr and [M/H] = -0.7 dex at $dm$ = 14.8 mag. To find this best combination, we explore an isochrone grid that covers a metallicity range of -2.2 $\leq$ [M/H] $\leq$ -0.2 and an Age range of 10 $\leq$ Age $\leq$ 13 with 0.1 dex and 1 Gyr spacing, along with a $dm$ varying from 13 to 16 with a step of 0.1 mag. The stream's strength is measured through the total weights between $-2\degr <$ offset $< 2\degr$ in Figure~\ref{fig:profile} and the best-fit result is found when the stream signals become the strongest. The PM term of \texttt{StreamScanner} is fixed at (-0.4, 0.8) and (-1, 1.4) as mentioned before during this process. After the PM selection and background subtraction, the stream's main sequence along with its turn-off is clearly seen and has a good match with the isochrone. The $dm$ = 14.8 mag corresponds to a heliocentric distance of $\sim$ 9.12 kpc. Considering the width of 1.9$\degr$, the physical width of Yangtze is about 302 pc, implying a dwarf galaxy origin. 

In the right panel of Figure~\ref{fig:CMD_PMs}, we present a 2D histogram of PMs. Similarly, before the subtraction between the stream region and the mean of the off-stream regions, a CMD selection is applied to them as shown with the red polygon in the left panel. The diagram with bin size = 0.2 mas/yr is also smoothed using a 2D Gaussian with $\sigma$ = 1 pixel. An overdensity at $\mu^*_{\alpha} \sim$ -0.7 mas/yr and $\mu_{\delta} \sim$ 1.1 mas/yr is discernable corresponding to the stream.  

We further estimate the stream's surface density and brightness. There are a total of 296 stars within the PM polygon after the background subtraction. This serves as an estimate to the number of the stream stars located in a $27\degr \times 2\degr$ region. Thus the surface density is roughly 5.5 stars degree$^{-2}$. For all stars in the stream region, each one is assigned a weight by \texttt{StreamScanner} and this allows us to select the most likely members of the stream based on the sorting of weights. We adopt stars with weights $>$ 1.255 as the member candidates because the criterion leaves us 296 stars as well. By combining their individual $G$ magnitudes, we get a surface brightness of Yangtze to be $\Sigma_G \simeq$ 34.85 mag arcsec$^{-2}$. We can put this another way. We have 298 stars satisfying the CMD and PM selections simultaneously, which can be treated as the member candidates as well. Those stars give a surface brightness of $\Sigma_G \simeq$ 34.95 mag arcsec$^{-2}$, very close to the previous one. 

\subsection{Association with GCs and Streams}

We aim to fit an orbit to Yangtze such that we can investigate whether it is related to any GCs or known streams of the Milky Way. We adopt a Galactic potential model used in \citet{2022A&A...667A..37Y,2022MNRAS.513..853Y}, where the halo is taken from \citet{2017MNRAS.465...76M} and non-halo components come from \citet[model I]{2017A&A...598A..66P}. The position of the sun is set as $(R_{\sun}, Z_{\sun})$ = (8.122, 0.0208) kpc \citep{2018A&A...615L..15G,2019MNRAS.482.1417B}, and the solar velocities are set to $(V_{R,\sun}, V_{\phi,\sun}, V_{Z,\sun})$ = (-12.9, 245.6, 7.78) km s$^{-1}$ \citep{2018RNAAS...2..210D}, respectively. The fitting parameters are position $\alpha$, $\delta$, heliocentric distance $d$, PMs $\mu^*_{\alpha}$, $\mu_{\delta}$, and radial velocity $V_r$. We chose to anchor the declination at $\delta$ = 0$\degr$, near the midpoint of the stream, leaving other parameters free to be varied. In a Bayesian framework, sky positions and PMs of 298 member candidates are used to constrain the parameters and the fitted results can be derived from their marginalized posterior distributions through a Markov Chain Monte Carlo sampling. We have cross-matched these candidates with spectroscopic catalogs like LAMOST or SDSS, but unfortunately no common stars are found thereby no radial velocities available. The best-fit parameters are $\alpha = 146.14^{+0.06}_{-0.06}$ $\degr$, $d = 8.58^{+0.39}_{-0.53}$ kpc, $\mu^*_{\alpha} = -0.76^{+0.02}_{-0.02}$ mas yr$^{-1}$, $\mu_{\delta} = 0.96^{+0.02}_{-0.01}$ mas yr$^{-1}$ and $V_r = 8.78^{+5.16}_{-5.03}$ km s$^{-1}$. We then obtain the stream's orbit by integrating it for $\pm$ 1 Gyr under the same potential.  

\begin{figure*}
	\includegraphics[width=\linewidth]{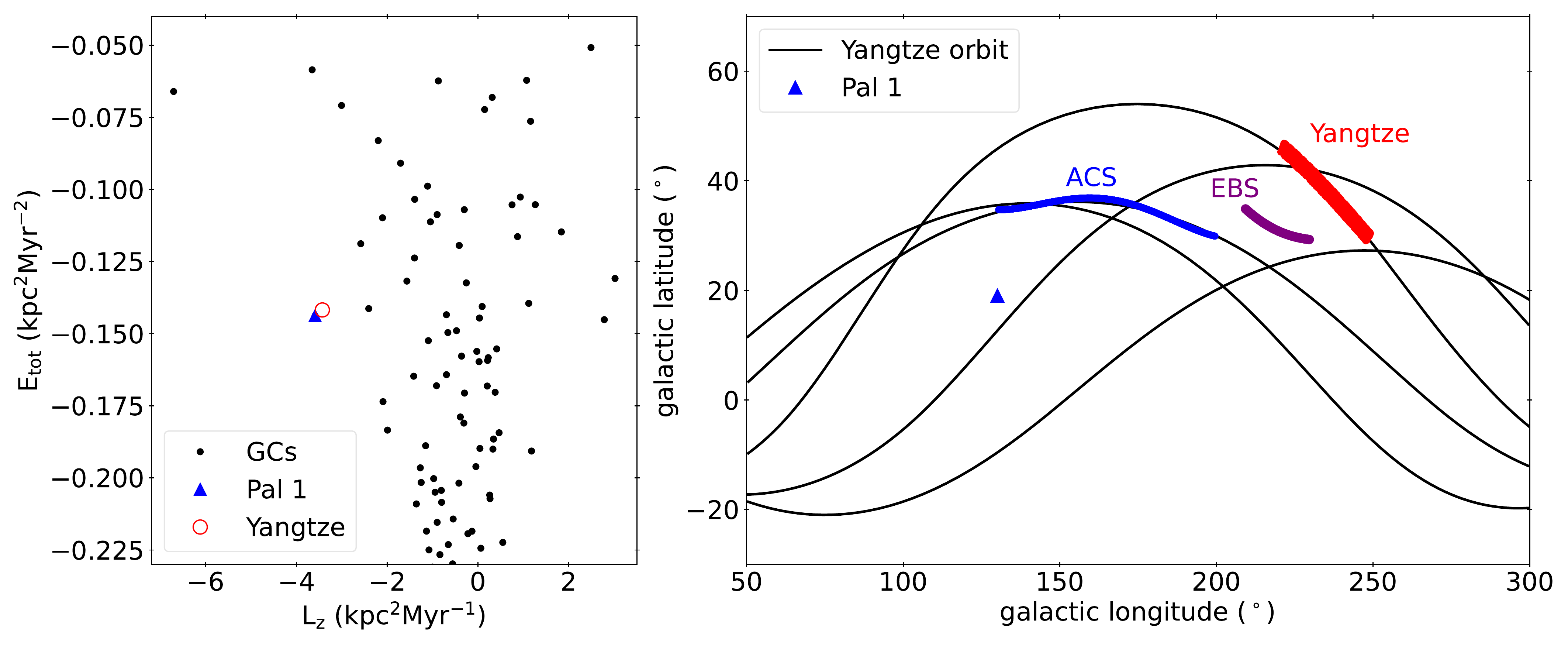}
	\caption{The left panel displays GCs (black points) and Yangtze (red circle) in angular momenta and energy space. The right panel presents projections of streams and orbits in Galactic coordinate. The red track and black line indicate Yangtze and its best-fit orbit. The blue and purple lines represent ACS and EBS, respectively. The blue triangle denotes Pal 1. 
		\label{fig:dynamics}}
\end{figure*}

We first examine possible connections between Yangtze and GCs by comparing their angular momenta $L_z$ and energy $E_{\rm tot}$ as shown in the left panel of Figure~\ref{fig:dynamics}. The positions and velocities of GCs are taken from \citet{2021MNRAS.505.5978V}. We note that the stream lies rather near Palomar 1 (Pal 1) as marked by the blue triangle. \citet{2011ApJ...740..106S} already showed that Pal has unusual chemical characteristics including lower [$\alpha$/Fe] ratios than Galactic stars of the same [Fe/H], and they concluded an extragalactic origin for Pal 1. The metallicity of Pal 1 is -0.6 dex which is close to -0.7 dex of Yangtze although the latter is estimated by an isochrone fitting. However, Yangtze's orbit does not pass through Pal 1 as displayed in the right panel of Figure~\ref{fig:dynamics}, and we have also verified that the cluster's orbit does not match with the stream's track either. Thus it might be a better interpretation that Pal 1 is a GC brought in by the progenitor of Yangtze. 

Among known streams, we find that the trajectory of the Anticenter Stream \citep[ACS;][]{2006ApJ...651L..29G} matches quite well with the orbit of Yangtze as shown in the right panel of Figure~\ref{fig:dynamics}. Here the footprint of ACS is based on results of \citet{2021A&A...646A..99R}, which is extracted from \texttt{galstreams} library \citep{2022arXiv220410326M}. ACS is likely on the Yangtze's orbit though they are separated by one orbit around the Galaxy. Yangtze's pericenter and apocenter are $R_{peri}$ = 12.1 kpc and $R_{apo}$ = 18.9 kpc, close to $R_{peri}$ = 15.4 kpc and $R_{apo}$ = 19.0 kpc of ACS \citep{2008ApJ...689L.117G}. It is also worth noting that our metallicity estimate is nearly identical to [Fe/H] = -0.72 dex of ACS \citep{2022ApJ...933..151Z}. All similarities connect these two streams together and imply that they may share a common origin. Besides ACS, we also show the Eastern Banded Structure \citep[EBS;][]{2011ApJ...738...98G}, another substructure in the anticenter region. The three streams are located rather close and nearly parallel to one another on sky but for EBS and Yangtze, their inferred orbits suggest that they are dynamically distinct. 

\section{Conclusion} \label{sec:summary}

With revised photometry and astrometry from $Gaia$ DR3, we show a new stream probably formed from a dwarf galaxy's disruption, which we dub Yangtze. The stream is detected at a significance of $\sim 15\sigma$ with a \texttt{StreamScanner} method that assigns weights to stars in CMD and PMs simultaneously. The Yangtze is spanning 27$\degr$ by 1.9$\degr$ on sky at 9.12 kpc away from the sun. Its stellar population is well fitted with an isochrone of Age = 11 Gyr and [M/H] = -0.7 dex. The stream has a low surface brightness of $\Sigma_G \simeq$ 34.9 mag arcsec$^{-2}$, with a density of about 5.5 stars degree$^{-2}$. We also demonstrate that Yangtze may be associated with ACS and Pal 1, and we suspect that these two streams are remnants of one dwarf galaxy that brought the cluster into the Milky Way. Spectroscopic observations will be required to give deeper insights to it.

\begin{acknowledgments}
	We thank the referee for the thorough reviews that helped us to improve the manuscript. This study is supported by the National Natural Science Foundation of China under grant nos 11988101, 12273055, 11973048, 11927804, 11890694, 11873052 and 12261141689, and the National Key R\&D Program of China, grant no. 2019YFA0405500. This work is also supported by the GHfund A (202202018107). We acknowledge the support from the 2m Chinese Space Station Telescope project CMS-CSST-2021-B05. 
	
	This work presents results from the European Space Agency (ESA) space mission Gaia. Gaia data are being processed by the Gaia Data Processing and Analysis Consortium (DPAC). Funding for the DPAC is provided by national institutions, in particular the institutions participating in the Gaia MultiLateral Agreement (MLA). The Gaia mission website is \url{https://www.cosmos.esa.int/gaia}. The Gaia archive website is \url{https://archives.esac.esa.int/gaia}.
	
\end{acknowledgments}





\bibliography{sample631}{}
\bibliographystyle{aasjournal}



\end{document}